\documentclass[fleqn,twoside]{article}
\usepackage{espcrc2}
\def\lesssim{\lower.7ex\hbox{${\buildrel < \over \sim}$}}
\def\gtrsim{\lower.7ex\hbox{${\buildrel > \over \sim}$}}
\def\mp{\lower.7ex\hbox{${\buildrel - \over +}$}}

\usepackage{graphicx}
\title{Charge ratio of muons from atmospheric neutrinos}
\author{T.K. Gaisser \& Todor Stanev\\
Bartol Research Institute, University of Delaware\\
Newark, DE 19716 USA}
\begin{document}
\begin{abstract}
We calculate the intensities and angular distributions of
positive and negative muons produced by atmospheric neutrinos.
We comment on some sources of uncertainty in the charge
ratio.  We also draw attention to a potentially interesting signature 
of neutrino oscillations in the muon
charge ratio, and we discuss the prospects
for its observation (which are not quite within the reach of
currently planned magnetized detectors).
\vspace{1pc}
\end{abstract}
\maketitle

The MINOS detector~\cite{Minos} under construction in the Soudan Mine
for study of neutrino oscillations with a neutrino beam from 
Fermilab will be able to
measure the momentum and charge of muons in the energy
range from just below one GeV to about 70~GeV/c.~\cite{Michael}
In addition to its use as the far detector for accelerator
neutrinos, MINOS is sufficiently large to detect a significant
number of atmospheric neutrinos. 
  Muons entering from outside the detector as well as muons
 originating from vertices inside the detector will occur.
 These events will provide a useful calibration beam for
 the detector.  

The detector, which is now under construction at Soudan, has already
detected atmospheric muons and neutrinos.  After completion, MINOS
will continue to collect cosmic-ray data for some time
before the accelerator neutrino beam turns on and afterwards as well.  
Because the
atmospheric neutrino beam has comparable fluxes of neutrinos and
anti-neutrinos, it will be possible~\cite{letter,Michael} 
to search for any sign of differences in oscillation properties of
$\nu_\mu$ and $\bar{\nu}_\mu$.  Such differences could arise in theories
in which there is intrinsic violation of CPT~\cite{CPT}.
They could also arise from matter
effects, for example with mixing of three active neutrino
flavors if $\sin^2\theta_{12}$ is large
enough~\cite{Bernabeu} or with non-maximal 
$\nu_\mu\leftrightarrow\nu_s$ mixing.~\cite{Ronga}

The purpose of this letter is to calculate the rates of
atmospheric neutrino interactions in a detector like MINOS
as a function of direction and energy separately for 
neutrinos and anti-neutrinos.  We briefly discuss some sources
of systematic uncertainty, and we note a potentially interesting effect
of oscillations as reflected in the charge ratio of atmospheric
muons.

To illustrate the basic result, we start with the atmospheric
neutrino spectrum of Ref.~\cite{AGLS}.
 We integrate the spectra
 of atmospheric neutrinos folded with $d\sigma/dy$ to obtain the
rate of events with contained vertices.  To obtain the rates of
external muons, the convolution also includes the muon
 range.~\cite{G&S84,G&S85}
 For simplicity
 we use the GRV94 structure functions~\cite{GRV94} to obtain the
 spectra of neutrino-induced $\mu^+$ and $\mu^-$.  (We have checked that
 the results for the limited range of $x$ and $Q^2$ relevant here
 are essentially the same as those obtained with more complicated
 structure functions.)  
In Table 1 we show expected event rates
as a function of muon energy with and without oscillations, separately
for contained vertices and for external upward-moving neutrino-induced muons.
Full mixing with $\Delta m^2=2.5\times 10^{-3}$~eV$^2$~\cite{Mauger}
was assumed for the oscillation case.  Contained vertices are
shown integrated over all directions; neutrino-induced muons
are summed over the upward $2\pi$~steradian.

\begin{table}[thb]
\caption{ Event rates for vertex contained and upward external neutrino
 induced muons at the location of MINOS for the epoch of solar minimum
 in four energy intervals, with and without oscillations.
 The oscillation parameters used are \protect$\sin^2{2 \theta}$ = 1
 and \protect$\Delta m^2$ = 0.0025~eV\protect$^2$.
 The units for vertex contained events are \protect$10^{-16} g^{-1}.s^{-1}$
 and for external upward going muons are \protect$10^{-13} cm^{-2}.s^{-1}$ 
 \label {tab_rates} }
\vspace*{3truemm}
\begin{tabular}{|r |c c|c c|} \hline
 E, GeV &\multicolumn{2}{|c|}{no oscillations}&\multicolumn{2}{|c|}{\protect$\Delta m^2$= 0.0025 eV\protect$^2$} \\ \hline
 &\protect$ \mu^-$ & \protect$ \mu^+$ & \protect$ \mu^-$ & \protect$ \mu^+$   
\\ \hline
 & \multicolumn{4}{|c|}{contained vertex} \\
  1 - 5      & 6.88 & 3.27 & 5.14 & 2.45 \\
  5 - 10     & 0.86 & 0.41 & 0.66 & 0.31 \\
 10 - 20     & 0.48 & 0.24 & 0.38 & 0.18 \\
\protect$>$20& 0.49 & 0.23 & 0.46 & 0.21 \\
\hline
 & \multicolumn{4}{|c|}{external upward going} \\
  1 - 5      & 3.57 & 1.71 & 2.01 & 0.94 \\
  5 - 10     & 1.58 & 0.75 & 1.05 & 0.48 \\
 10 - 20     & 1.55 & 0.72 & 1.21 & 0.54 \\
\protect$>20$&5.38 & 2.19 &  5.20 & 2.10 \\
\hline
\end{tabular}
\end{table}

The first point to note is that the charge ratio for neutrino-induced
muons is reversed relative to that for atmospheric muons.
There are more negative than positive neutrino-induced muons
($\mu^+/\mu^-\lesssim 1/2$),
whereas for muons produced in the atmosphere, 
$\mu^+/\mu^-\approx 1.25$~\cite{atmosmu}
This is a consequence of the ratio of cross sections,
$\sigma_\nu/\sigma_{\bar{\nu}}>1$, and the fact that 
$\nu_\mu\rightarrow\mu^-$ while $\bar{\nu}_\mu\rightarrow \mu^+$.
In addition, the ratio $\nu_\mu/\bar{\nu}_\mu$ increases slowly above
its low energy value of unity, first as muon-decay becomes unimportant
and at higher energy because of the increasing importance of the
channel $p\rightarrow \Lambda\,K^+\rightarrow \nu_\mu\,\mu^+$.

 Secondly, the energy spectrum of the external muons is significantly
harder than for the contained vertices.  This is a consequence
of the increase of the muon range with energy.  A much higher range
of neutrino energies contributes to the external events than
to the events with interaction vertex inside the detector.
For $E_\nu>100$~GeV kaons become the dominant source of neutrinos,
with a big contribution from the process
$p\rightarrow \Lambda\,K^+\rightarrow \nu_\mu$.  As a result, the
charge asymmetry is somewhat larger for external events than for
contained vertices.

%%%%
\begin{figure}[htb]
\includegraphics[width=220pt]{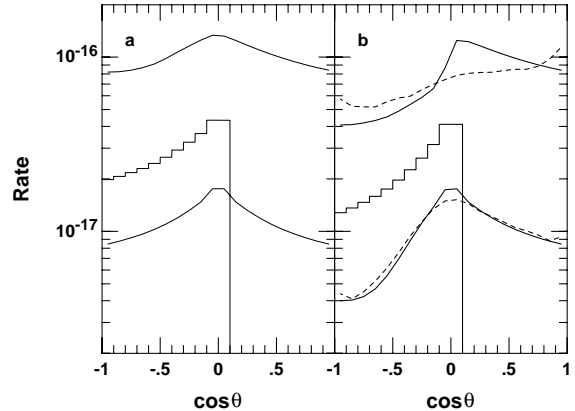}
\vspace*{-5truemm}
\caption{
 a) Zenith angle dependence of the neutrinos generating vertex contained
 muons of energy above 1 GeV (upper line) and 10 GeV (lower line)
 and of upward going external muons (histogram) in the absence of
 oscillations.
 The units for vertex contained events are \protect$(g.s.sr)^{-1}$
 and for external upward going muons are \protect$(cm^2.s.sr)^{-1}$. 
 The flux of upward going muons is divided by 10\protect$^4$.
  Note that the upward going muons are plotted up to \protect$\cos{\theta}$
 = 0.1; b) Same in the presence of oscillations with
  \protect$\sin^2{2 \theta}$ = 1
 and \protect$\Delta m^2$ = 0.0025. The dashed  lines show the 
 zenith angle dependence of the muons.
 \label{ang_dis}
}
\end{figure}

\begin{figure}[thb]
\includegraphics[width=220pt]{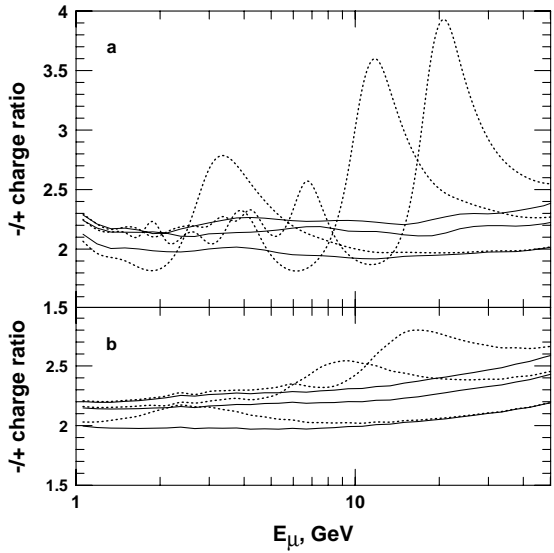}
\vspace*{-5truemm}
\caption{Charge ratio of neutrino-induced muons
 a) for events with contained vertices;
b) for external muons.  Solid lines are with no oscillations and
dotted lines with oscillations for three bins of zenith angle
centered on \protect$\cos\theta = -0.95$, $-0.55$ and $-0.15$.
Near vertical is highest solid line (rightmost dotted peak).
\label{charge_rat}
}
\end{figure}

 It is also interesting to look at the angular dependence of
 neutrino-induced muons with and without oscillations. 
 Fig.~1 shows the zenith angle dependence of the
 neutrinos that produce muons with contained vertices and upward going
 neutrino-induced muons originating outside the detector.
 The strongest effect of oscillations is on the contained
 vertex events above 1 GeV. 

 The symmetric shape of the no oscillation
 case is replaced by an asymmetric distribution with a 
deficit due to oscillations for upward muons from neutrinos with
pathlengths comparable to the radius of the Earth.
 This effect is in reality obscured by
 the large angle between the neutrino and muon direction 
 at low energy as
 shown by the dashed lines in Fig.~1b.
 The effect of the oscillations is also obvious in the higher energy
 ($E>10$ GeV) sample. The rate of vertically
 upgoing contained vertex muons is decreased by a factor
 of $\sim2$. The muon--neutrino angle in this case is significantly
 smaller and does not affect much the oscillation features.
 The smallest effect is in the case of external events. On one
 hand, the contribution of high energy, non-oscillating neutrinos
 is the largest and on the other the downgoing external muon
 rate is dominated by atmospheric muons, so most
of the downward hemisphere is not available as a probe of
neutrino oscillations with external events. The muon--neutrino angle 
 for this sample is negligible.
The background of atmospheric muons is low enough at
the depth of Soudan so that neutrino-induced, external
muons can be measured up to $\cos\theta\approx0.1$, about $6^\circ$
above the horizon.  Integrated over all upward-going external 
events the, oscillations decrease the rate by 22\%, which 
is comparable to the uncertainty in the expected flux.~\cite{GHLS}
Instead, the effect of oscillations is visible as a distortion of 
the angular distribution, as observed at MACRO~\cite{MACRO} and
Super-K.~\cite{SKup}

 We assume here that neutrinos and anti-neutrinos have identical
 oscillations properties.  Because of
 the properties of the differential charged current cross sections
of neutrinos,
however, the relation between $E_\nu$ and $E_\mu$ is different for
 $\nu_\mu$ and $\bar{\nu}_\mu$.  As a consequence, when the 
muon charge ratio is plotted as a function of muon energy
for a given angular bin (corresponding to a given pathlength)
oscillation features appear in the ratio because the 
oscillation minima show up at different energies
for $\mu^+$ and $\mu^-$.  Figure~2 illustrates this for
contained vertices (a) and for external muons (b).
The ratios are shown for three
 bins of $\cos\theta$ 
 going from nearly vertical ($-1< \cos\theta < -0.9$) to 
nearly horizontal ($-0.2<\cos\theta<-0.1$).
 The solid lines show the charge ratios in the absence of
 oscillations. The baseline charge ratio increases slightly
at high energy and is larger near the vertical because of
the increased importance of kaon decay relative to pion decay.

Calculation of external events involves an
integral over all contributing neutrino energies taking account
of the increase of muon range with energy.  
 The growth of the charge ratio with muon
 energy is stronger than for contained vertices
because of the contribution of higher energy
 neutrinos. In the oscillation case the variations are
 smaller than for contained vertex events and do not show the
 secondary sinusoidal features. For the external events, the peaks
 are wider and shallower than for contained vertices because
of the contribution of a broader range of neutrino energies.

%%%%%%%%%%%%%%%%%%%%%%%%%%%%%%%
\begin{figure}[h]
\includegraphics[width=220pt]{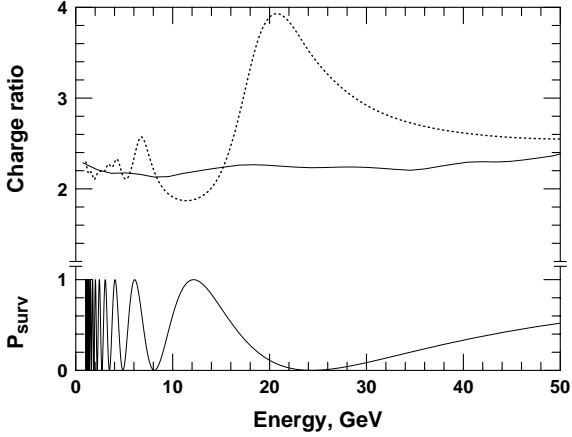} 
\vspace*{-5truemm}
\caption{
 Upper part shows the variation of the \protect$\mu^-/\mu^+$ charge
 ratio of almost vertical (\protect$\cos{\theta}$ = -0.95), vertex
 contained upward-going neutrino-induced muons as a function of
 muon energy for no oscillations (solid) and oscillations (dotted line).
 The lower part shows the neutrino
 survival probability for the same angular bin. 
 \label{surv_prob}}
\end{figure}
%%%%%%%%%%%%%%%%%%%%%%%%%%%%%%%%%%

 For more insight into the origin of this behavior, we show in Fig.~3
 the charge ratio with and without oscillations for the vertically
 upward bin of zenith angle, for which the pathlength $L\sim 10^4$~km.
 The lower panel shows the corresponding survival probability, which
 is assumed identical for neutrinos and antineutrinos.  An
 approximate, quantitative understanding of the features
 can be obtained by considering
 the leading term in the differential neutrino
 charged current cross sections.  For neutrinos
\begin{equation}
{d\sigma(E_\nu)\over dz^-}\;\approx\; \sigma_0\times E_\nu/GeV
\label{neutrino}
\end{equation}
 and 
\begin{equation}
{d\sigma(E_{\bar{\nu}})\over dz^+}\;\approx\; \sigma_0\times 
(z^+)^2\times E_{\bar{\nu}}/GeV
\label{antineutrino}
\end{equation}
for antineutrinos.  Here $0\le z = E_\mu/E_\nu\le 1$
and $\sigma_0\approx 0.8\times 10^{-38}$cm$^2$.
Using the approximation~\ref{neutrino} we can
estimate the relation for neutrinos
by evaluating the average $E_\nu$ for the distribution
\begin{equation}
\phi(E_\nu){d\sigma(E_\nu)\over d E_{\mu^-}}.
\label{weight}
\end{equation}
Here 
\begin{equation}
\phi(E_\nu)\;\propto\;E_\nu^{-(\gamma+1)}
\label{power}
\end{equation}
is a power-law approximation for the differential spectrum of neutrinos.
Then 
\begin{equation}
\langle E_\nu\rangle\;\approx\;{\gamma\over\gamma-1}E_{\mu^-}.
\label{mumin}
\end{equation}
The corresponding result for antineutrinos is
\begin{equation}
\langle E_{\bar{\nu}}\rangle\;\approx\;
{\gamma+2\over\gamma+1}E_{\mu^+}.
\label{muplus}
\end{equation}

 With $\gamma\approx 2$ in the relevant energy range~\cite{AGLS},
 we have $E_{\bar{\nu}}/E_{\mu^+}\approx 4/3$ and $E_\nu/E_{\mu^-}\approx2$.  
 Thus the first oscillation minimum at $\approx 24$~GeV
should be reflected
 at $\approx 18$~GeV in $\bar{\nu}_\mu$ and $\mu^+$ as an
 increase in the $\mu^-/\mu^+$ ratio and at $\approx 12$~GeV in
 the opposite channel, reflected as the minimum in the
 $\mu^-/\mu^+$ ratio below 12~GeV. The slight deviations from these
 numbers in Fig.~3 reflect the breadth and asymmetry of the oscillation
features folded with the neutrino spectrum.

  Unfortunately, the MINOS detector is not large enough to
 see the oscillation effect in the charge ratio well, if at all.
 For example, we estimate that there will be only some
 12 $\mu^-$ and 5 $\mu^+$ events with energy between 10 and 30 GeV 
 in 25~kt-yr in the angular region $-1<\cos\theta<-0.5$,
 where the charge ratio is high (see Fig.~2). Our estimates are based on
 the angular distribution of the interacting neutrinos and do not
 account for the experimental efficiency.

\begin{figure}[htb]
\includegraphics[width=220pt]{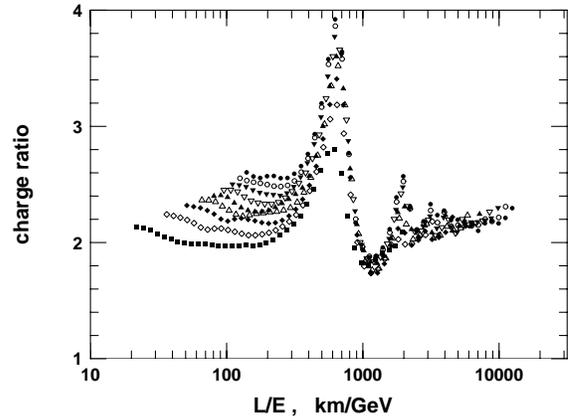} 
\vspace*{-5truemm}
\caption{
 Charge ratio of \protect$\mu^-/\mu^+$ for upward going contained
 vertex neutrino induced muons as a function of L/E for different
 zenith angles.
 Different symbols show events in \protect$\cos{\theta}$ bins
 of -0.15, -0.25, -0.35, ... , -0.95. The horizontal case is not
 plotted because of the very strong change of the oscillation length
 which requires  a much more exact calculation.
 \label{chargeLE}
}
\end{figure}

 The best way to use all the data is to bin every event by $L/E$.
 Fig.~4 shows the charge ratio in upward going contained vertex
 events for the standard oscillation parameters as a function of
 $L/E$ for nine different $\cos{\theta}$ bins. The horizontal
 bin is not included because the fast change of the oscillation
 length there requires a much more exact calculation.
 One can easily identify the peaks and valleys in the charge
 ratio that follow the oscillation structure of the
 atmospheric neutrino flux. The main feature is at $L/E$ of
 $\sim$~600, where the charge ratio can grow to almost twice its
 baseline value of $\mu^-/\mu^+\approx 2$.
 The valley is at $L/E$ of $\sim$~1200. 
 One can subdivide the $L/E$ range in two parts with different
 charge ratios, 2.36 in the $L/E$ from 300 to 900 and 1.84 from 900
 to 1500. The expected number of muons for 25 kT.yr are
 26.7 $\mu^-$ \& 11.3 $\mu^+$ and 15.5 \& 8.4 respectively, still
 very small statistics. 

 One should also note that Fig.~4 shows the neutrino pathlength $L$
 which is not experimentally measured. For some of these events,
 where the sinusoidal variation happens at energies below 10 GeV,
 the angular smearing of the neutrino--muon angle will seriously
 obscure the effect when plotted as $L_\mu/E_\mu$.

 It is interesting to ask if one could distinguish oscillations
 from neutrino decay with an observation of the muon charge ratios.
 The $L/E\; <$ 1500 part of Fig.~4 would not change for a decay
 scenario that fits the muon disappearance observations. At higher
 $L/E$ values the charge ratio behavior would be much smoother
 without the secondary and tertiary peaks at $L/E$ of about 2000
 and 3000. The increase of the charge ratio at these positions are
 however small and even more difficult to detect.

 The reliable detection of the variation of the muon charge ratio
 would require a much bigger detector. The MONOLITH~\cite{MONOLITH}
 detector with mass of 30 kT was proposed for the Gran Sasso
 Laboratory. If this detector had been built and were
 operated for five years it would be able to measure the effect
 at the 3$\sigma$ level.

 The MINOS detector will be able to measure the total neutrino
 energy but assume that the muon and neutrino directions coincide.
 The difference between the neutrino and muon energy spectra as a
 function of L, as well as the charge ratios, could provide additional
 observable parameters for the more exact derivation of the
 oscillation parameters. 

 The exact manifestation of the effect however has systematic
 uncertainties related to the uncertainties of the atmospheric
 flux. These arise from the limited knowledge of the primary
 cosmic ray spectrum and the production of pions and kaons. 
 The overall uncertainty  for $E_\nu\sim10$~GeV is estimated in
 Ref.~\cite{GH} as $\pm 25$\%.

 The uncertainty in the ratio $\nu_\mu/\bar{\nu}_\mu$ 
 should be significantly less than the uncertainty in the magnitude
 of the flux because the normalization of the primary cosmic ray
 spectrum cancels. The largest remaining source of uncertainty
 in the ratio is from production of charged pions and kaons,
 especially in the forward fragmentation region, as reflected by
 the spectrum weighted moments, $Z_{p\pi^+}$, $Z_{p\pi^-}$, $Z_{pK^+}$
 and $Z_{pK^-}$.  For pions and especially for kaons, the positive
 channel dominates, and kaons become relatively more important at
 high energy.

  We can estimate the uncertainty in the $\nu_\mu/\bar{\nu}_\mu$ 
 ratio by combining in quadrature the estimates in Ref.~\cite{AGLS}
 of uncertainty in the flux of $\nu_\mu + \bar{\nu}_\mu$ from each
 of the four channels independently.  This leads to an estimate of
 $\pm8$\% below $10$~GeV and $\pm9$\% for $10 < E_\nu < 100$~GeV.
 Analysis of data from forthcoming hadron production experiments with
 CERN~\cite{HARP,NA49} and Fermilab~\cite{E907} should lead to a 
 reduction of this estimate.

 The effect we described above should also be detectable in a slightly
 different form in the long baseline oscillation experiment with the NuMI 
 neutrino beam. The experimental statistics would be fully sufficient
 and the exact manifestation would strongly depend on the beam
 energy spectrum, purity, and composition. All of these three 
 parameters are currently still at the design stage.
 
%% {\bf Discuss increasing importance of kaons and its implication for
%% estimating the uncertainty the $\nu_\mu/\bar{\nu}_\mu$ ratio.}
%% For this we need response curves for $>1$GeV contained, $>10$GeV contained
%% and external showing separately the pion and kaon contributions.
%% An example is Fig.~5 of my semianalytic response paper (Astroparticle
%% Physics 16, 2002).

\noindent
{\bf Acknowledgments.}  We are grateful for helpful discussions with Doug Michael
and Teresa Montaruli.  This work is supported in part by the U.S. Dept.
of Energy under Grant No. DE-FG02-91ER40626.

\end{document}